\documentclass{elsart1p}
\usepackage{graphicx}
\usepackage{amssymb}
\usepackage{subfigure}
\usepackage{tabls}
\usepackage{amssymb,amsmath}
\usepackage{setspace} 

\newcommand{\pt}{\ensuremath{p_{\mathrm{t}}}}

\renewcommand{\{}{\left\{}
\renewcommand{\}}{\right\}}

\begin{document}
%
\begin{frontmatter}
%
%
%
%
%
\title{The development of high performance online tracker for High Level Trigger of Muon Spectrometer of ALICE}
%
%

\author{Indranil Das (for ALICE Collaboration)}
\ead{Indranil.Das@cern.ch}
\address{High Energy Nuclear and Particle Physics Division, Saha Institute of Nuclear Physics, \\1/AF Bidhannagar, Kolkata-700064, India}

\begin{abstract}

The Muon Spectrometer (MS) of the ALICE experiment at LHC is equipped
with a HLT (High Level Trigger), whose aim is to improve the accuracy
of the trigger cuts delivered at the L0 stage. A computational
challenge of real-time event reconstruction is satisfied to achieve
this software trigger cut of the HLT. After the description of the
online algorithms, the performance of the online tracker is compared
with that of the offline tracker using the measured pp collisions at
$\sqrt{s}=7$ TeV.    


\end{abstract}

\end{frontmatter}

\section{Introduction}

Lattice calculations of Quantum ChromoDynamics (QCD), predict that at a
critical temperature ($\sim$ 170 MeV) and energy
density ($\sim$ 1 GeV/fm$^3$), the nuclear matter undergoes a phase
transition to a deconfined state of quarks and gluons~\cite{LQCD}, also known as
Quark Gluon Plasma (QGP). In the laboratory, the QCD phase diagram
and the expected formation of the QGP can be studied using ultra-relativistic heavy-ion collisions. The
suppression~\cite{Ch1_Matsui-Satz} or enhancement~\cite{Ch1_pbm_js} of heavy quark resonances in 
heavy-ion collisions at LHC is predicted to be a strong signature of
QGP as the temperature of the
fireball produced in Pb-Pb collisions is expected to be about
three times higher than the critical temperature.

ALICE (A Large Ion Collider  
Experiment~\cite{Ch1_PPR1,Ch1_PPR2,JINST_PPR}) is a general purpose experiment
whose detectors identify and measure hadrons, electrons, photons and
muons produced in p-p and Pb-Pb collisions at the CERN LHC. ALICE is
optimized for heavy-ion reactions and thus is capable to track and
identify particles from very low ($\sim$ 100 MeV/c) to fairly high
($\sim$ 100 GeV/c) transverse momentum, to reconstruct short lived
particles like hyperons, D and B mesons and the heavy quark resonances [J/$\psi$ 
($c\bar{c}$), $\Upsilon$ ($b\bar{b}$)] in an environment of extreme
particle densities. The J/$\psi$ and $\Upsilon$ are reconstructed by ALICE in
the central rapidity region ($-0.9<y<0.9$) from their dielectron
decay channel and at large rapidity ($2.5<y<4.0$) from their dimuon
decay channel using a spectrometer.

\begin{figure}[ht]
\centering
\subfigure{
\includegraphics[height=7.0cm,width=10.0cm]{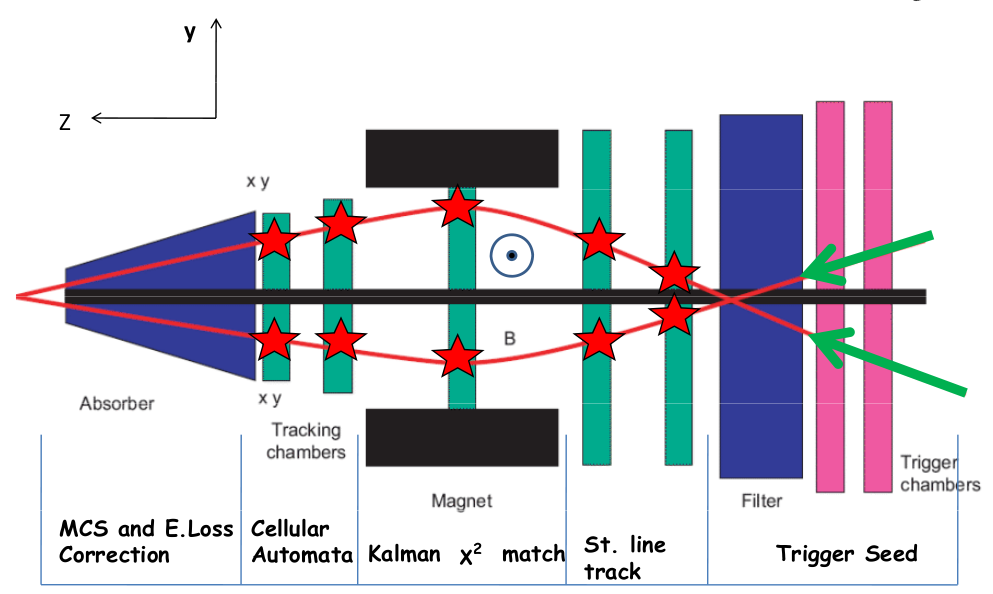}
}
\hspace{0.1cm}
\vspace{0.1cm}
\subfigure{ 
  \includegraphics[height=7.0cm,width=10.0cm]{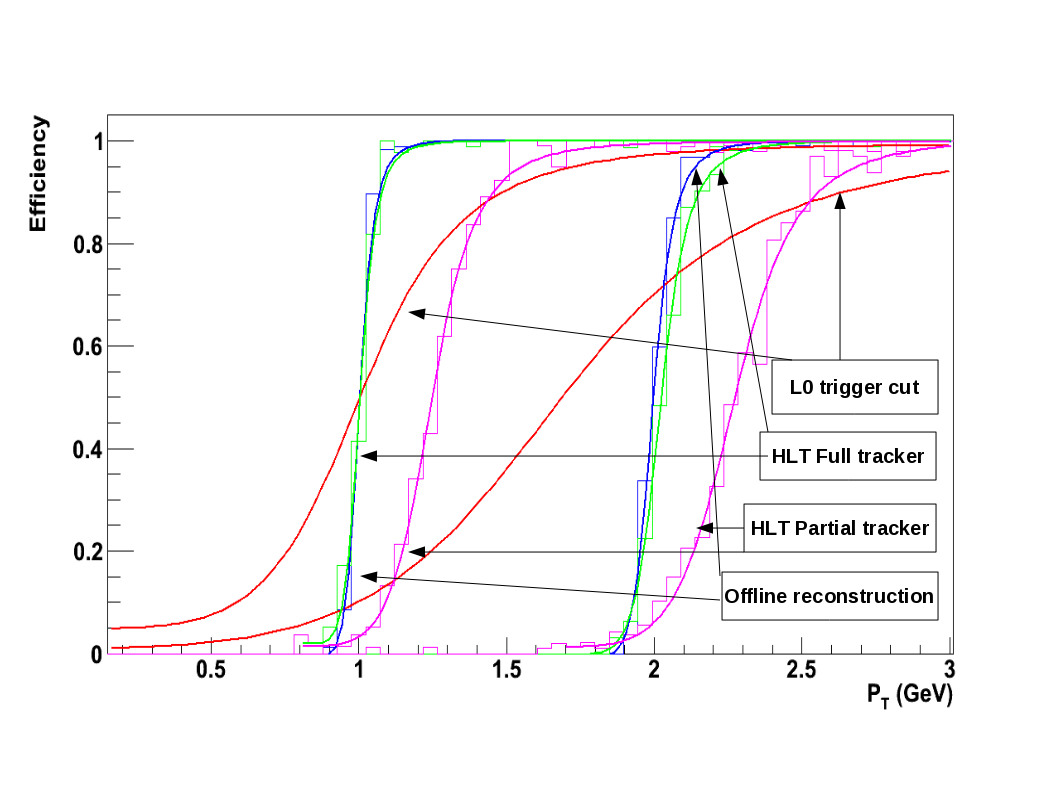}
}
\caption{\label{fig:MS_Pt_Cut_Efficiency} The two dimensional y-z plane view of the
  MS (top) and the efficiencies of transverse momentum
  ($p_{\mathrm{T}}$) cuts calculated for various
  reconstruction methods (bottom).}  
\end{figure}

This Muon Spectrometer [Fig.~\ref{fig:MS_Pt_Cut_Efficiency}
(top)] is designed to run at the highest dimuon rate in heavy-ion
collisions at LHC ($\mathcal{L}\sim10^{27}$~cm$^{-2}$~s$^{-1}$ for
Pb-Pb beam). 
It consists of the following components: a passive front
absorber to absorb hadrons and photons from the interaction vertex;
high granularity tracking system of 5 stations each with two detection
planes; a large warm dipole magnet; a passive muon filter wall,
followed by four planes of trigger chambers and a inner beam shield
surrounding the beam pipe to protect the chambers from high particle
flux at large rapidities.  

Each tracking chamber provides two dimensional hit information by
measuring the charge distributions on two segmented cathode
planes. The cathode which has higher resolution in the direction
(y-direction) perpendicular to the plane containing the magnetic field
(x-direction) and beam axis (z-direction) is referred to as the
bending cathode, while the other along the magnetic field is referred to as
non-bending cathode. Two tracking stations are placed before, one
inside and two after the dipole magnet. The total number of readout
pads is about 1.1 million and covers an area of about 100 m$^2$. The
trigger system consists of four Resistive Plate Chamber (RPC) planes
arranged in two stations which are placed behind the muon filter. The
trigger system has to select events containing a muon pair coming from
the decay of J/$\psi$ or $\Upsilon$ resonances from all possible
background contaminations \footnote{However, the trigger system also
  provides single muon triggers to identify events with single muon
  track.}. The main background 
comes from low-\pt~ muons of pion and kaon decays. Thus, the L0 trigger is
generated if at least two tracks with opposite charge (or the same charge used in
this case for background subtraction), both above a predefined \pt~ cut
are detected in an event. Two different \pt~ thresholds of 1 GeV/c and 2
GeV/c have been chosen for J/$\psi$ and $\Upsilon$ measurement,
respectively, according to simulation studies. 

However, the coarse grained segmentation of the RPCs and the presence 
of the iron wall before the trigger stations, do not allow a sharp \pt~
cut. This is demonstrated in Fig.~\ref{fig:MS_Pt_Cut_Efficiency}
(bottom), through a simulation study carried out using the AliRoot
framework~\cite{Offline}. It is evident that the L0 trigger passes a
substantial number of events which are below the \pt-threshold. In
addition, the trigger efficiency is around 60\% near the threshold
value of 1 GeV/c and 2 GeV/c. Thus, the primary task of the dimuon
High Level Trigger (dHLT) is to refine the \pt~ cut in order to
increase its selectivity. For this purpose, in the 
addition to the trigger chambers, the slower but more accurate tracking
chambers have been used and a real-time event reconstruction scheme
for the entire Muon Spectrometer has been developed. 

This scheme should satisfy the following criteria:

(a) The event processing should be done at a rate of 1 kHz for Pb-Pb
collisions; (b) The real-time reconstruction should produce 
results of appreciable quality without losing any signal event; (c) It
should be robust and should not stop processing due to any data
corruption in the input buffer.

The event reconstruction in tracking chambers is a two step process,
which involves the reconstruction of charge clusters and the track
formation using the reconstructed hit points. In offline
reconstruction, these two steps are implemented by Mathieson fitting of the charge
clusters which are defined by the nearest neighbour algorithm and
Kalman filtering, respectively. However, none of them can be applied for
real-time reconstruction due to the time constrain set by the expected
event rate of 1 kHz. Thus, new algorithms for fast reconstruction have been
developed for dHLT~\cite{BB_HLT}. These are described below.

\section{Algorithms}

\vspace{0.5cm}
\begin{description}
\item [ \textbf{ Hit-Reconstruction: }]  This new algorithm does not
  identify the charge clusters by nearest neighbour search. Instead, it
  searches for pads with maximum charge whose immediate neighbours
  have nonzero charge. The two schemes are equivalent since every
  cluster has an unique central pad whose charge is greater 
  than other members. Once the central pads are identified, the hit
  positions along the bending and non-bending directions are given by
  the centre of gravity of the charges measured on the central pad and
  the two pads around it in y and x directions, respectively. Finally
  the information on bending (y-direction) and non-bending
  (x-direction) cathodes are merged to generate hit position.

\item [ \textbf{ Partial Tracking: }] In the first attempt to improve
  the accuracy of the L0 trigger, the trigger track segments are
  extended upto the fourth tracking
  station~\cite{BB_HLT,Ch4_MansoTracker}. The straight line tracks 
  in the two tracking stations after the magnetic field give a better
  \pt~ estimation than the value obtained from the trigger tracks. The
  result of this improvement is shown in
  Fig.~\ref{fig:MS_Pt_Cut_Efficiency} (bottom), by the solid lines
  which are marked by `HLT Partial Tracker'. It is evident that there
  is a marked improvement in accuracy over the L0 trigger, but the
  trigger efficiencies at \pt~ = 1 GeV/c and 2 GeV/c are about 20\% and
  the desired efficiency ($>$90\%) is achieved only around 1.5 GeV
  and 2.5 GeV, respectively. Thus, this method is not suitable for
  dHLT and the full tracking through the magnetic field is essential. 

\item [ \textbf{ Full Tracking: }] The Full Tracking formalism
  creates tracks from the last trigger station to
  the first tracking station. Various tracking stages has been shown
  in  Fig.~\ref{fig:MS_Pt_Cut_Efficiency}(top). These are described
  below. 

  \begin{description}
  \item [{\it Tracking in Station 4,5:}] This part is identical to
    the Partial Tracking method where the straight line tracks  are
    extrapolated upto the fourth tracking station.
  \item [{\it Tracking in Station 1,2:}]  In this part, the small
    track segments are formed in station 1 and 2 
    using Cellular Automata~\cite{APPP_A_CA-Kisel-1} (CA) method. The
    CA formalism has been followed since it does not require any a
    priori knowledge of the seed for the tracking. 
  \item [{\it Kalman $\chi^2$-test:}]  Once the track segments are
    formed before and after the dipole magnet, they are matched
    through the magnetic field using $\chi^2$ test of Kalman
    filtering~\cite{APPP_B_Fruehwirth}. If the track
    segments are not matched, the \pt~ of the track is estimated as in
    the case of Partial Tracking. This ensures that no physics event
    is lost in the Full Tracking scheme. 
  \item [{\it Track Extrapolation:}] In case of complete tracks, the
    \pt~ estimation is further improved by incorporating the corrections due
    to energy loss and multiple Coulomb scattering in the front
    absorber. 
  \end{description}

\end{description}
 
\section{Results}

The accuracy and efficiency of the \pt-cut for the Full Tracker is
shown in Fig.~\ref{fig:MS_Pt_Cut_Efficiency} (bottom), by the solid
line marked as `HLT Full Tracker' and have been compared with the
offline reconstruction. It can be observed from this simulation study that
the estimated \pt's of the muon tracks match closely with the offline
analysis and the desired improvement of the L0 can be achieved.  

The framework of the Full Tracker was successfully implemented on the
HLT computing farm and all the validation tests were completed before
the pp collisions at $\sqrt{s}=7$~TeV data taking in March, 2010.

\begin{figure}[ht]
\centering
\subfigure{
\includegraphics[width=10cm,height=7cm]{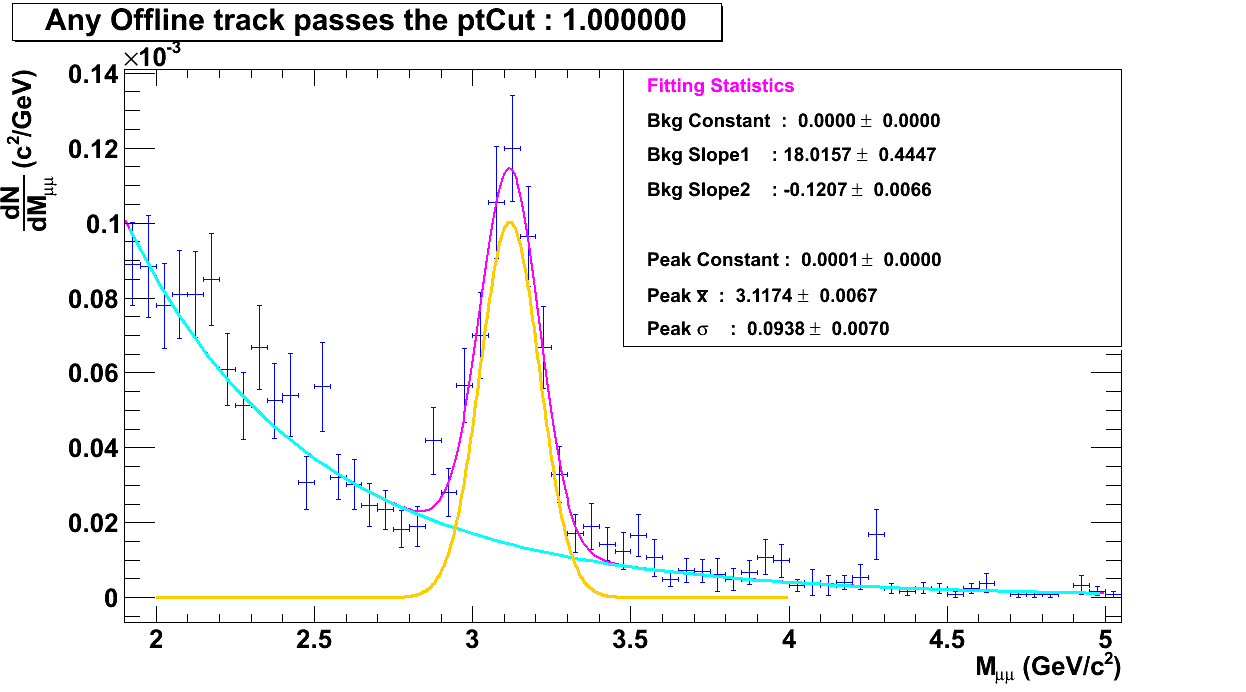}
}
\hspace{0.1cm}
\vspace{0.1cm}
\subfigure{
\includegraphics[width=10cm,height=7cm]{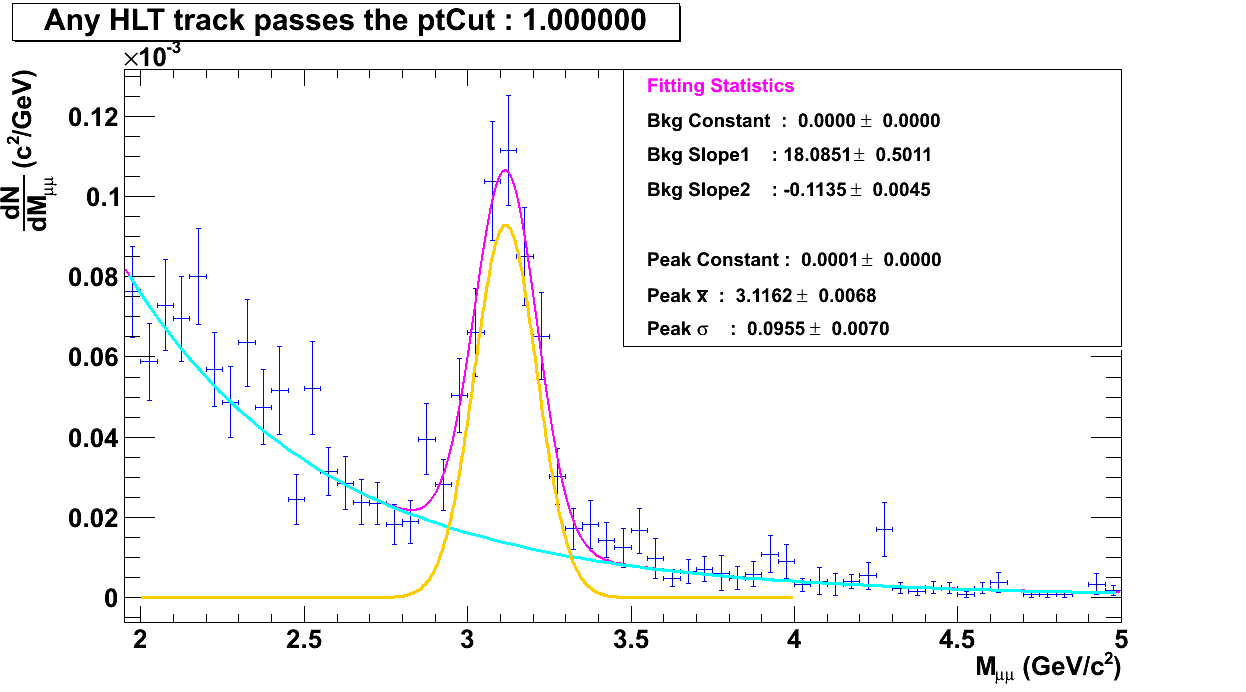}
}
\caption{\label{fig:Offline_HLT_Trigger_Cut} The event averaged invariant mass
  spectrum with a $p_{\mathrm{T}}$-cut of 1 GeV from offline
  reconstruction (top) and Full Tracker (bottom). See the text for
  more detail.}  
\end{figure}

In Fig.~\ref{fig:Offline_HLT_Trigger_Cut}, the effect of the \pt-cut
from the offline analysis and from the real-time Full Tracker has been
compared for pp (``real data'') collisions. The top and bottom panels of the
figure show the invariant mass plot around the J/$\psi$ peak when the
\pt~ of one muon was found to be greater than 1 GeV/c by the offline
and online reconstruction, respectively. The invariant spectra are
fitted with a  Gaussian signal and a  double exponential functions. The width of the
J/$\psi$ peak was found to
be 93.8 and 95.5 MeV/c$^2$ for offline and HLT triggers,
respectively. This shows that a sufficient accuracy of the
$p_{\mathrm{T}}$-cut for the Full Tracker is achieved. However, the counts in the
J/$\psi$ peak for the offline reconstruction is 
higher by 7\% with respect to the online reconstruction. This indicate a
small inefficiency of the online trigger. Studies are ongoing to
correct this effect.

\section{Summary}

It has been demonstrated both from simulation and data, that the HLT
for the Muon Spectrometer is capable to improve the trigger momentum
resolution of the tracks and validate the L0 muon candidates by a
sharp \pt~ cut. In the coming years of LHC operation, when the
luminosity of the beams will reach its nominal values, the dHLT will
play a crucial role in background rejection and allow the Muon
Spectrometer to run at higher L0 trigger rates. In addition, it is
estimated that the use of dHLT, with a loosening of the L0 \pt~ cut,
will allow recovering about 20\% of the low-\pt~ muon events. 


  

\end{document}